\pdfoutput=1
\documentclass[preprint2,times]{aastex6}

\usepackage{natbib}
\usepackage{amsmath}
\usepackage{amssymb}
\usepackage{url}

\renewcommand{\vec}[1]{ {\mathbf #1} }

\newcommand{\Fig}{{Figure}}
\newcommand{\Figs}{{Figures}}

\providecommand{\dodoi}[1]{doi:~\href{http://doi.org/#1}{\nolinkurl{#1}}}
\providecommand{\url}[1]{\href{#1}{#1}}
\providecommand{\doeprint}[1]{\href{http://ascl.net/#1}{\nolinkurl{http://ascl.net/#1}}}
\providecommand{\doarXiv}[1]{\href{https://arxiv.org/abs/#1}{\nolinkurl{https://arxiv.org/abs/#1}}}

\begin{document}

\title{Formation of Magnetic Flux Rope during Solar Eruption. I. \\
Evolution of Toroidal Flux and Reconnection Flux}

\author{Chaowei Jiang\altaffilmark{1}, Jun Chen\altaffilmark{2,3}, Aiying Duan\altaffilmark{4}, Xinkai Bian\altaffilmark{1}, Xinyi Wang\altaffilmark{5}, Jiaying Li\altaffilmark{1}, Peng Zou\altaffilmark{1}, Xueshang Feng\altaffilmark{1,5}}

\altaffiltext{1}{Institute of Space Science and Applied Technology, Harbin Institute of Technology, Shenzhen 518055, China}
\altaffiltext{2}{School of Astronomy and Space Science, Nanjing University, Nanjing 210023, China}
\altaffiltext{3}{CAS Key Laboratory of Geospace Environment, School of Earth and Space Sciences, University of Science and Technology of China, Hefei, Anhui 230026, China}
\altaffiltext{4}{School of Atmospheric Sciences, Sun Yat-Sen University, Zhuhai 519000, China}
\altaffiltext{5}{State Key Laboratory for Space Weather, Center for Space Science and Applied Research, Chinese Academy of Sciences, Beijing 100190, China}

\email{*Corresponding author: chaowei@hit.edu.cn}

\begin{abstract}
  Magnetic flux ropes (MFRs) constitute the core structure of coronal
  mass ejections (CMEs), but hot debates remain on whether the MFR
  forms before or during solar eruptions. Furthermore, how flare
  reconnection shapes the erupting MFR is still elusive in three
  dimensions. Here we studied a new MHD simulation of CME initiation
  by tether-cutting magnetic reconnection in a single magnetic
  arcade. The simulation follows the whole life, including the birth
  and subsequent evolution, of an MFR during eruption. In the early
  phase, the MFR is partially separated from its ambient field by a
  magnetic quasi-separatrix layer (QSL) that has a double-J shaped
  footprint on the bottom surface. With the ongoing of the
  reconnection, the arms of the two J-shaped footprints continually
  separate from each other, and the hooks of the J shaped footprints
  expand and eventually become closed almost at the eruption peak
  time, and thereafter the MFR is fully separated from the
  un-reconnected field by the QSL. We further studied the evolution of
  the toroidal flux in the MFR and compared it with that of the
  reconnected flux. Our simulation reproduced an evolution pattern of
  increase-to-decrease of the toroidal flux, which is reported
  recently in observations of variations in flare ribbons and
  transient coronal dimming. The increase of toroidal flux is owing to
  the flare reconnection in the early phase that transforms the
  sheared arcade to twisted field lines, while its decrease is a
  result of reconnection between field lines in the interior of the
  MFR in the later phase.
\end{abstract}

\keywords{Magnetic fields; Magnetohydrodynamics (MHD); Methods: numerical; Sun: corona; Sun: flares}

\section{Introduction}
\label{sec:intro}

Solar eruptions are spectacular manifestation of explosive release of
magnetic energy in the Sun's atmosphere, i.e., the solar corona, and
therefore, unveiling the relevant magnetic field structures and their
evolution holds a central position in the study of solar
eruptions. Magnetic flux rope (MFR), a bundle of twisted magnetic
field lines winding around a common axis with the same sign, is
believed to be a fundamental structure in solar
eruptions~\citep{ChengX2017, DuanA2019, LiuR2020}, especially in those
which successfully produce coronal mass ejections (CMEs). By
reconstruction of the cross section of ICME (i.e., CMEs that evolves
into the inter-planetary space) from the in-situ data obtained by
satellites passing through the ICME, it has been well established that
typical ICMEs have structure of highly twisted
MFR~\citep[e.g.,][]{WangY2016, HuQ2017}.

Although there is little doubt that MFR constitutes the core structure
of CMEs, whether MFR exists in the solar corona before CME initiation
is still in intense debates~\citep{ChenP2011,
  Patsourakos2020}. Currently, there are two different opinions; one
is that MFR does not exist before solar eruption, and it is the latter
that creates MFR through magnetic reconnection; the other is that MFR
should exist prior to eruption and it is the ideal magnetohydrodynamic
(MHD) instability of the MFR that initiates the eruption. The typical
scenarios for the first opinion include the runaway tether-cutting
reconnection model~\citep{Moore1980, Moore1992, Moore2001} and the
magnetic breakout model~\citep{Antiochos1999, Aulanier2000, Lynch2008,
  Wyper2017}. In these models, the coronal magnetic field before
eruption is strongly sheared and eruption is triggered by magnetic
reconnection, internally within the sheared arcade (i.e.,
tether-cutting), or above it (i.e., breakout), while MFR is built up
during the eruption through reconnection which transforms the sheared
arcade into the rope. For the models assuming the pre-existence of
MFR, such as the catastrophe model~\citep{Forbes1991, LinJ2001}, the
torus instability and kink instability models~\citep{Kliem2006,
  Torok2005, Fan2007, Aulanier2010, Amari2018}, an MFR is proposed to
either emerge from below the photosphere~\citep[i.e., in the
convective zone, where the turbulent convection can create thin,
twisted magnetic tubes, ][]{Fan2001, Cheung2014}, or forms slowly by
reconnection in the lower atmosphere~\citep{Green2011} through the
so-called flux cancellation
process~\citep{Ballegooijen1989}. Observations seem to indicate that
both opinions are possible. For example, on one hand,
\citet{SongH2014} presented a good observation that an MFR can formed
during a CME. On the other hand, an MFR characterized by a hot sigmoid
structure may pre-exist before eruption, as manifested by precursor
oscillation~\citep{ZhouG2016} or precursor external magnetic
reconnection between the top of the MFR and ambient magnetic
field~\citep{ZhouG2019}.

Regardless of which model is relevant to the real case in the corona,
it is commonly agreed that flare reconnection (i.e., the main
reconnection that occurs below the erupting MFR) can shape
substantially the on-the-fly MFR. In the purely two-dimensional (2D)
standard flare model, a plasmoid (corresponding to the cross section
of MFR in 3D) rises from the top of the flare current sheet, and
reconnection in the current sheet continuously adds poloidal flux to
the MFR, which thus grows and expands during the eruption. As a
result, the observed double flare ribbons, which indicate the
locations of the reconnecting field-line footpoints in the opposite
magnetic polarities, are continuously separated with each
other. However, in a fully 3D case, it is not that straightforward how
the reconnection shapes the MFR. Numerical simulations of the simplest
magnetic configuration (i.e., a bipolar magnetic field), aided with
accurate analysis of magnetic topology evolution, have been developed
to study how an MFR evolves with reconnection during
eruption~\citep{Aulanier2010, Aulanier2012}, and the findings are
becoming known as the standard flare model in 3D~\citep{Aulanier2012,
  Janvier2013, Janvier2014}, although it is still an over-simplified
version of the realistic cases as demonstrated in recent
data-constrained and data-driven
simulations~\citep[e.g.,][]{JiangC2018, JiangC2018a, ZhongZ2021}.

In the standard 3D model,
the erupting MFR is separated from the ambient field by a
quasi-separatrix layer~\citep[QSL][]{Demoulin1996}, and in more
details, this QSL intersects with itself below the MFR, forming a
hyperbolic flux tube~\citep[HFT, ][]{Titov2002}, and the flare
reconnection occurs mainly in the HFT. The footprints of this QSL at
the bottom surface, i.e., the photosphere, forms two thin strips of J
shape on each side of the main polarity inversion line (PIL), and the
legs of the MFR are anchored within the hooked parts of the J-shaped
strips. Thus, the observed flare ribbons usually exhibit double-J
pattern, and the transient coronal holes~\citep{Kahler2001}, i.e.,
post-eruptive twin coronal dimmings, are naturally suggested to map
the feet of erupting MFRs, along which mass leakage into
interplanetary space could take place~\citep{Webb2000, Qiu2007,
  XingC2020}, and the boundaries of such twin coronal dimmings are
outlined by the hooks of flare ribbons. With the ongoing of
reconnection, the arms of the double-J ribbons separate, and their
hooks gradually extends outwards.  In such process, the flare
reconnection, which occurs between the pre-reconnection sheared arcade
\citep[as shown in the classic cartoon of tether-cutting model, i.e.,
Figure 1 of][]{Moore2001}, should increase the toroidal (axial) flux
by increasing the number of field lines within the MFR.

However, such a `standard' type of flare reconnection in 3D still
cannot explain fully the observations of `standard' two-ribbon
flares. A well-known, unexplained fact is that the feet of the
erupting flux rope, as manifested by twin coronal dimmings and also by
the hook ends of double-J flare ribbons, are found to be drifting
progressively away from the main PIL during
eruption~\citep{Kahler2001}, even though the photosphere can be
regarded as motionless during the short time scale of eruption. To
this end, \citet{Aulanier2019} analyzed in more details the
reconnection process in their simulation of flux rope eruption and
showed that the flare reconnection actually occurs in three different
types of events according to their different effect in building up the
flux rope. The first one is named as \emph{aa-rf reconnection}, which
is the standard 3D flare reconnection that occurs between two arcades
and results in a long field line joining the flux rope and a short one
as a flare loop. The second one is the so-called \emph{rr-rf
  reconnection}, which occurs within the flux rope by reconnecting two
flux-rope field lines with each other and generates a new multi-turn
flux-rope field line and a flare loop. The third one is \emph{ar-rf
  reconnection}, in which an inclined arcade reconnects with the leg
of a flux-rope field line, and it generates new flux-rope field line
rooted far away from the PIL and a flare loop. Thus, it is the
\emph{ar-rf reconnection} that actually leads to gradual drifting of
the MFR footpoints.

Observations show even more features not explained (or not mentioned)
in the `advanced' standard model of \citet{Aulanier2019}. For
instance, using high-resolution observations, \citet{WangW2017} found
two closed-ring-shaped flare ribbons in the case of a buildup of
highly twisted MFRs with the development of a flare
reconnection. During the separation of the main flare ribbons, the
flare rings expand significantly, starting from almost point-like
brightening.
Note that such closed circular shape flare ribbons have different
nature from those formed by null-point topology which also produce
circular ribbon due to reconnection in the null's spine-fan
separatrix~\citep{JiangC2018a}. It is predicted by theoretical
models~\citep{Demoulin1996} that if the MFR grows to sufficiently
twisted, the hooks of the double-J shaped footprints of the QSL can
indeed close onto themselves, becoming two closed rings, although this
is not reproduced by the numerical model of~\citep{Aulanier2010},
possibly because their simulation run is stopped before the MFR grows
to such a high degree of twist.

Very recently, a few papers reported that there is a systematic
decrease of the toroidal flux of erupting MFR after its fast
increase~\citep[e.g.,][]{WangW2017, ChenH2019, XingC2020}. In
particular, \citet{XingC2020} developed a practical method for
estimation of the toroidal flux of MFR during eruption by combining
twin coronal dimmings and the hooks of flare ribbons. They found that
the toroidal flux of the CME flux rope for all four studied
events shows a two-phase evolution: a rapid increasing phase followed
by a decreasing phase, and moreover, the evolution is well synchronous
in time with that of the flare soft X-ray flux. The increase of MFR's
toroidal flux can be easily understood by the \emph{aa-rf
  reconnection} while the subsequent decrease remains
unclear. Although \citet{XingC2020} invoked the \emph{rr-rf
  reconnection} as the mechanism responsible for such decrease, it is
still unknown whether the increase-to-decrease evolution of toroidal
flux can self-consistently be reproduced in any MHD simulation.

This series of papers are devoted to a comprehensive analysis of a new
MHD simulation of eruption~\citep{JiangC2021NA}, focusing on the
formation of MFR during eruption. That simulation demonstrated in
fully 3D that solar eruption can be initiated from a single magnetic
arcade without the formation of MFR before the triggering of
eruption. This is different from the aforementioned
simulations~\citep{Aulanier2010, Aulanier2019}, in which the eruption
is initiated by torus instability~\citep[which is a kind of ideal MHD
instability,][]{Kliem2006} of an MFR formed well before the
eruption. With the new MHD simulation, one can follow whole life,
i.e., the birth and subsequent evolution, of an MFR during the
eruption. As the first paper of this series, here we show, for the
first time, that both the closing of the hook ends of the QSLs at the
MFR's feet and the increase-to-decrease evolution of the toroidal
magnetic flux can be self-consistently reproduced by the simulation,
suggesting that they are genuine features of erupting MFRs. We further
quantified the evolution of reconnection flux during the eruption, and
found that the evolution of QSLs is rather complex within the MFR. In
the second article of this series, we will illustrate the 3D
configuration of the different types of magnetic reconnection in
building up the MFR and disclose why the QSLs evolve in such a complex
way, following the pioneering work by~\citet{Aulanier2019}.

\begin{figure*}[htbp]
  \centering
  \includegraphics[width=0.9\textwidth]{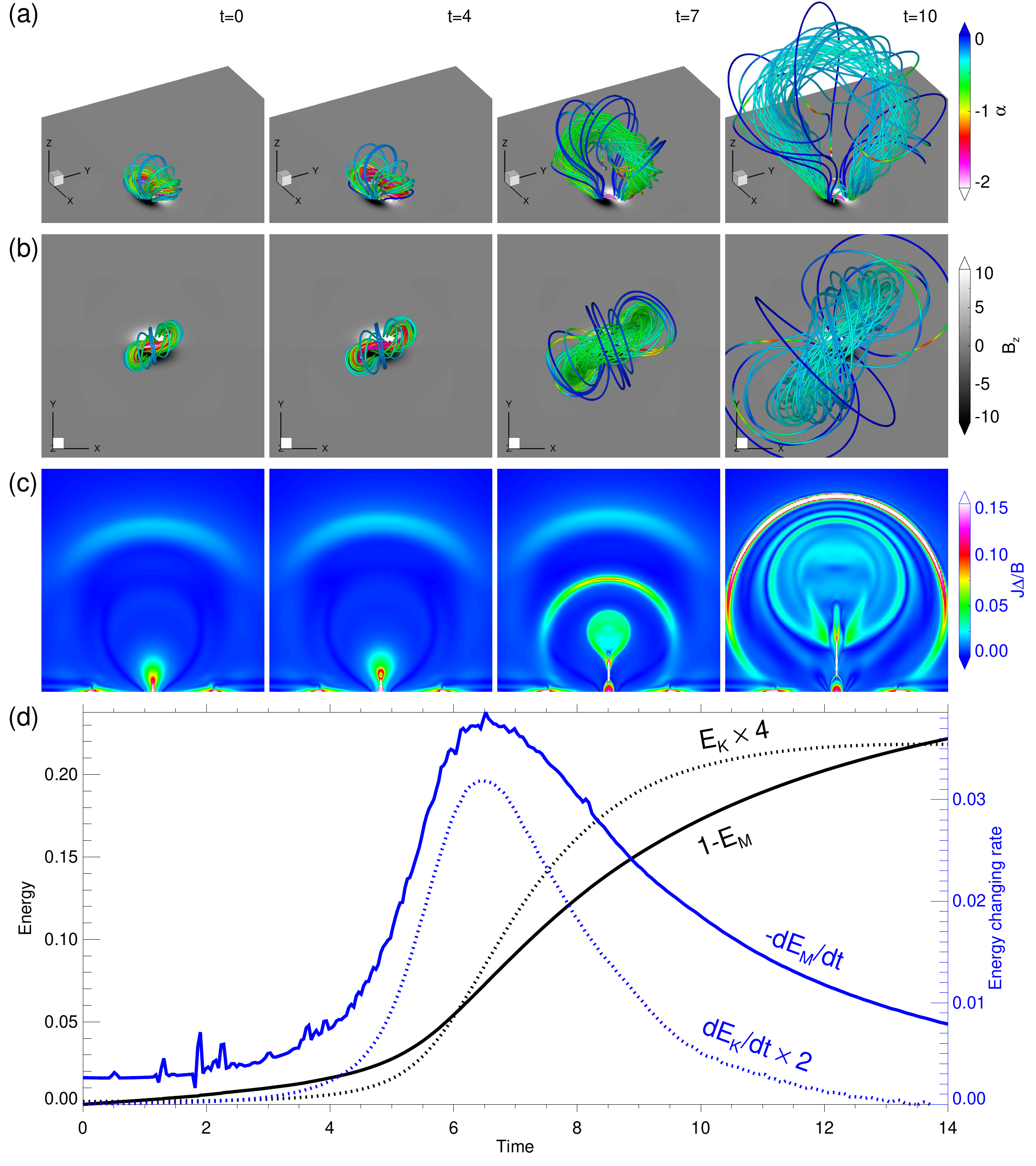}
  \caption{\textbf{Structural evolution of the eruption}. \textbf{(a)}
    3D prospective view of magnetic field lines colored by the
    force-free factor. Here the field lines are traced at fixed
    footpoints on the bottom surface, and they represent the core
    structure of the pre-eruption field. \textbf{(b)} Top view of the
    structure shown in (a). \textbf{(c)} Evolution of the
    dimensionless current density $J\Delta/B$ on the central cross
    section (i.e., the $x=0$ slice). \textbf{(d)} Evolution of
    magnetic and kinetic energies and their temporal changing
    rate. The energies are all normalized by the magnetic energy at
    $t=0$ and the unit of time is $105$~s. Also see Supplementary
    Movie 1 for a high-cadence evolution of the eruption process.}
  \label{eruption}
\end{figure*}

\section{MHD Simulation and Analysis Method}
\label{sec:method}

Recently, \citet{JiangC2021NA} performed an ultra-high accuracy, fully
3D MHD simulation and demonstrated that solar eruptions can be
initiated in a single sheared arcade with no additional special
topology. Their simulation shows that, ``through photospheric shearing
motion alone, an electric current sheet forms in the highly sheared
core field of the magnetic arcade during its quasi-static
evolution. Once magnetic reconnection sets in, the whole arcade is
expelled impulsively, forming a fast-expanding twisted flux rope with
a highly turbulent reconnecting region underneath''. They further
found that the high-speed reconnection jet plays the key role in
driving the eruption. The simplicity and efficacy of this scenario, in
the theoretical point of view, argue strongly for its fundamental
importance in the initiation of solar eruptions. Since the model do
not need a pre-existing MFR, the MFR itself comes into being
\emph{after} the eruption initiation.

Here we focus on the formation and evolution of MFR during the
eruption by using a simulation run like the one
in~\citet{JiangC2021NA}, but with a lower resolution than the original
ones. Such simulation solves the full set of MHD equations and starts
from a bipolar potential magnetic field and a hydrostatic plasma
stratified by solar gravity with typical coronal temperature. Then
shearing flows along the PIL, which are implemented by rotating the
two magnetic polarities at the photosphere in the same count-clockwise
direction, are applied on the bottom boundary to energize the coronal
field until an eruption is triggered, and after then the surface flow
is stopped. The whole computational box extends as
$(-32, -32, 0) < (x, y, z) < (32, 32, 64)$ with length unit of
$11.5$~Mm. We solve a full MHD equation with both solar gravity and
plasma pressure included, but with the energy equation simplified as
an isothermal process. The time unit of the model is $\tau = 105$~s,
and the shearing motion is applied by approximately $120 \tau$ before
the onset of the eruption, during which a current sheet is gradually
built up. Since no explicit resistivity is used in the MHD model,
magnetic reconnection is triggered when the current sheet is
sufficiently thin such that its width is close to the grid
resolution. For more details of the simulation settings, the readers
are referred to~\citet{JiangC2021NA}. In that paper, the simulation is
managed to be of very high resolutions with Lundquist number achieving
$\sim 10^5$ for a length unit. Therefore, the secondary tearing
instability (or plasmoid instability) is triggered in the current
sheet and the magnetic topology becomes extremely complicated in small
scales along with formation of the large-scale MFR. Such a complexity
substantially complicates our analysis of the large-scale magnetic
topology evolution associated with the erupting MFR, thus in this
paper we used a lower-resolution run (corresponding to a
  Lundquist number of $\sim 10^3$). In the lower-resolution run, the
basic evolution of the MFR during the eruption is not changed as
compared to the high-resolution run, except that the small-scale
structure will not arise, and thus the QSLs are computed in a cleaner
pattern. Moreover, with the lower resolution, we can run the
simulation longer and thus follow a longer evolution of MFR.

To help revealing the variation of the magnetic topology, we study the
distribution and evolution of two parameters, the magnetic squashing
degree and the magnetic twist number, which are commonly used for the
study of 3D magnetic fields and their dynamics~\citep{Aulanier2012,
  Janvier2013,Inoue2013,Savcheva2016, LiuR2016, DuanA2019}. The
magnetic squashing degree $Q$ quantifies the gradient of magnetic
field-line mapping with respect to their footpoints, and it is helpful
for searching QSLs (and true separatries) of magnetic
fields~\citep{Titov2002}, which can have extremely large values of $Q$
(e.g., $\geq 10^5$) and are preferential sites of magnetic
reconnection. By locating the QSLs from the high values of $Q$ we can
see how the magnetic topology is evolved by the magnetic
reconnection. Specifically, for a field line starting at one footpoint
$(x,y)$ and ending at the other footpoint $(X, Y)$ where $X$ and $Y$
are both functions of $x$ and $y$, the squashing degree $Q$ associated
with this field line is given by~\citep{Titov2002}
\begin{equation}
  \label{eq:Q}
  Q = \frac{a^{2}+b^{2}+c^{2}+d^{2}}{|ad-bc|}
\end{equation}
where
\begin{equation}
  a = \frac{\partial X}{\partial x},\ \
  b = \frac{\partial X}{\partial y},\ \
  c = \frac{\partial Y}{\partial x},\ \
  d = \frac{\partial Y}{\partial y}.
\end{equation}
The magnetic twist number $T_w$~\citep{Berger2006} is defined for a
given (closed) field line by taking integration of
$T_w=\int_L \vec J\cdot \vec B/B^2 dl /(4\pi)$ along the length $L$ of
the field line between two conjugated footpoints on the
photosphere. Note that $T_w$ is not identical to the classic winding
number of field lines about a common axis, but an approximation of the
number of turns that two infinitesimally close field lines wind about
each other~\citep{LiuR2016}.

\begin{figure*}[htbp]
  \centering
  \includegraphics[width=1\textwidth]{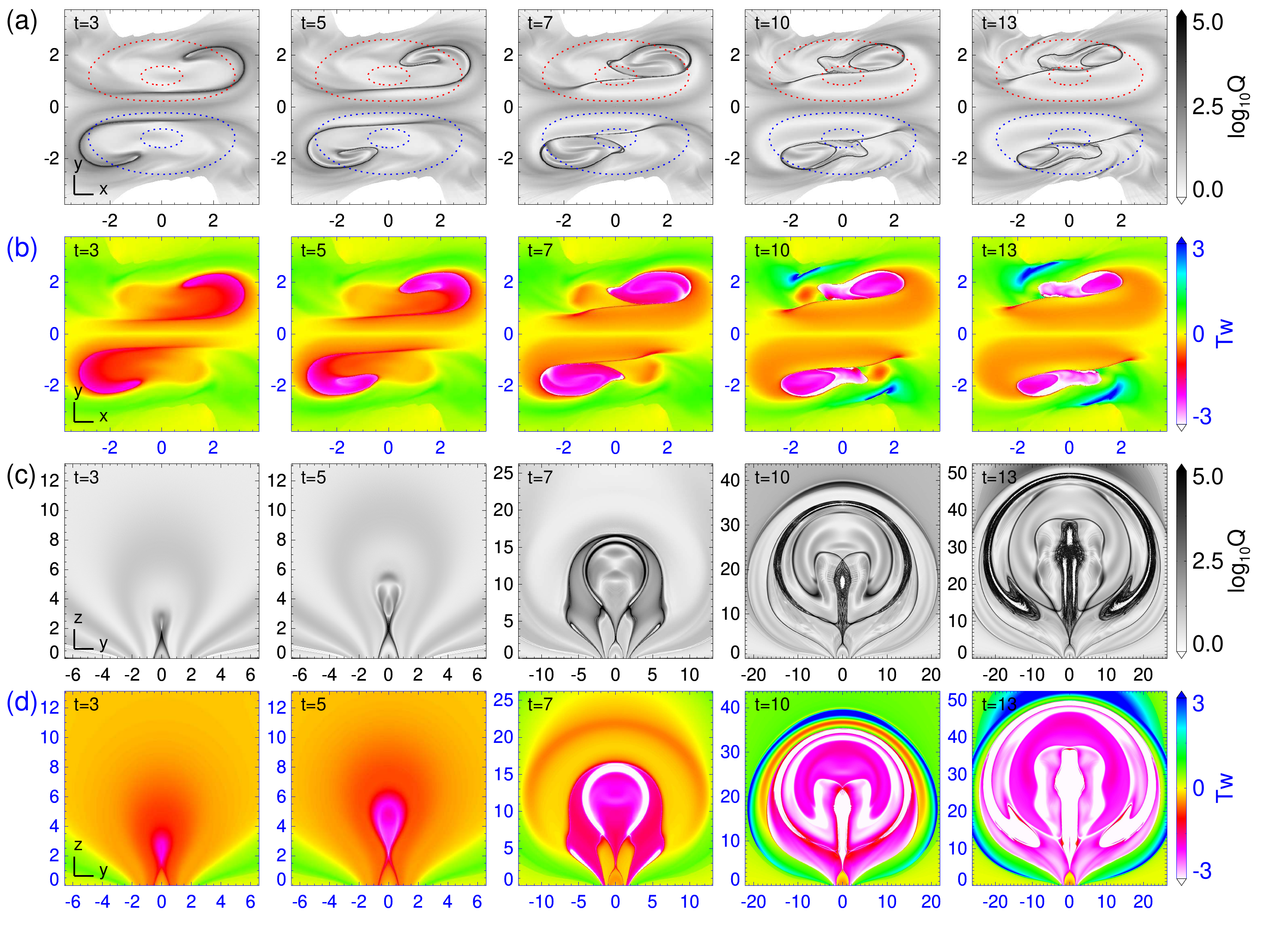}
  \caption{\textbf{Magnetic topology evolution and formation of MFR
      during the eruption.} \textbf{(a)} Magnetic squashing degree $Q$
    on the bottom surface. The dashed lines are contours of
    $B_{z}=-5$, $-10$ (blue) and $5$, $10$ (red). \textbf{(b)}
    Magnetic twist number $T_{w}$ on the bottom surface. \textbf{(c)}
    and \textbf{(d)} show the two parameters $Q$ and $T_{w}$ on the
    central vertical cross section (i.e., $x=0$ plane). Also see
    Supplementary Movie 2 for a high-cadence evolution.}
  \label{MFR_evol}
\end{figure*}

\begin{figure*}[htbp]
  \centering
  \includegraphics[width=0.8\textwidth]{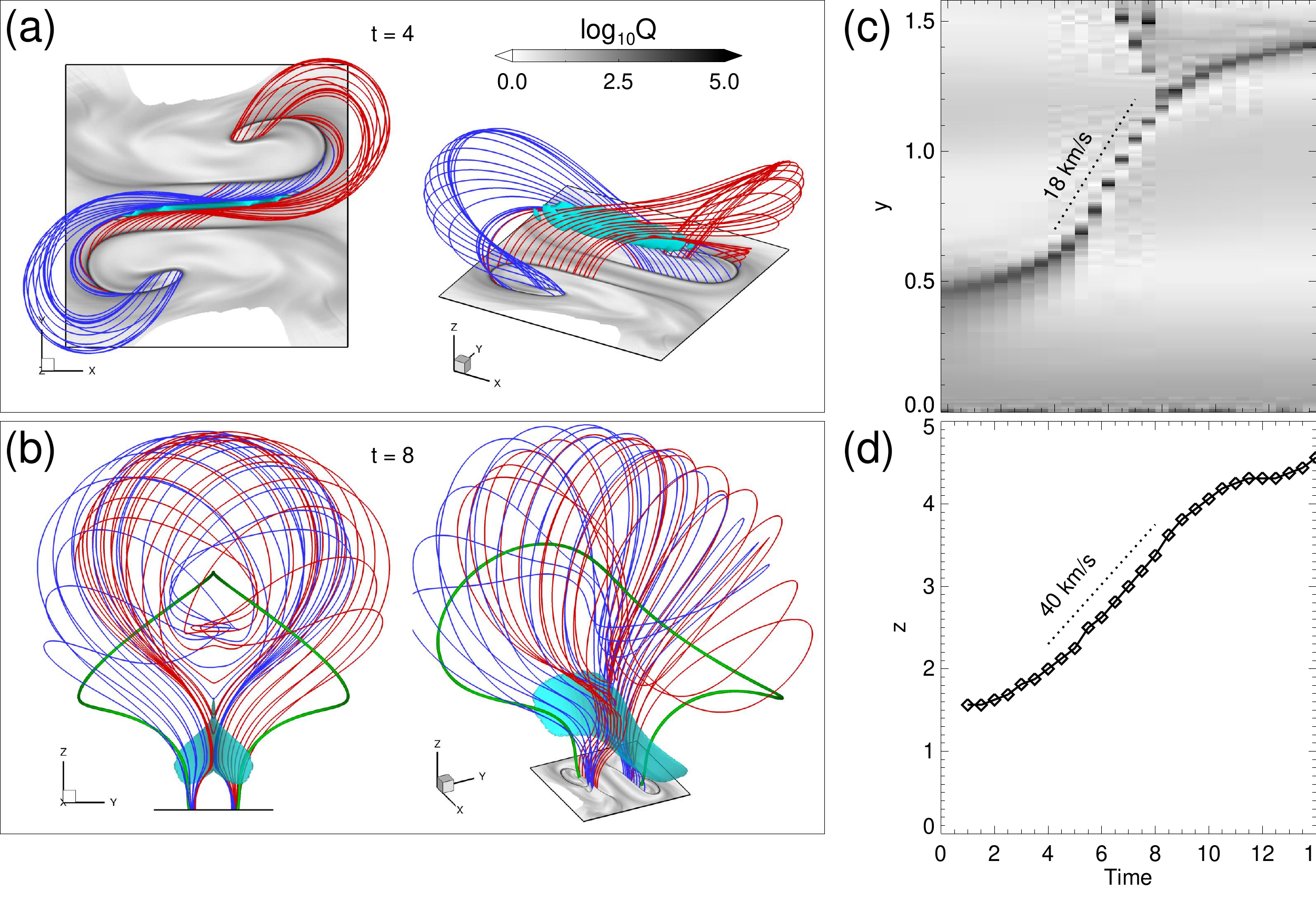}
  \caption{\textbf{Details of reconnection.} \textbf{(a)} Structure of
    the reconnecting field lines at $t=4$. The field lines are colored
    in red and blue. The cyan object is the iso-surface of
    $J\Delta/B=0.2$, i.e., the reconnection CS. The magnetic field
    lines are traced from the QSLs shown on the bottom surface, and as
    can be seen, they all contact the CS as these field lines are
    undergoing reconnection. \textbf{(b)} Sample of reconnecting field
    lines at $t=8$. The thick green lines represent the axis of the
    newly formed MFR. Note that the actual sizes of the bottom
    surfaces shown in (a) and (b) are identical. \textbf{(c)} A
    stacked time sequence of the bottom $Q$ map in $y$-direction and
    centred at $x=0$, which shows the separation motion of the two
    J-shaped QSLs shown in \Fig~\ref{MFR_evol}(a). The sloped, dashed
    line denotes the largest separation speed. \textbf{(d)} Time
    evolution of the height of X point of the hyperbolic flux tube,
    i.e., the apex of the cusp structure, shown in
    \Fig~\ref{MFR_evol}(c).  The slopped, dashed line denotes the
    largest rising speed.}
  \label{recon_sample}
\end{figure*}

\section{Results}
\label{sec:res}

\subsection{Overview of the eruption}
\Figs~\ref{eruption} (and Supplementary Movie~1) shows the magnetic
field lines, current density, and energies evolution from slightly
before the eruption onset to a time well after the eruption peak time
(that is, the peak time of energy conversion rate, which is
$t=6.5$). As can be seen, our simulation demonstrates a typical
coronal eruption leading to a CME, as seen in observations as well as
many previous numerical simulations with different
scenarios~\citep{Linker2003, Amari2003A, Torok2018}. The core magnetic
field changes from the pre-eruptive sheared arcades to a inverse
S-shaped MFR structure that subsequently exhibits a huge growth in
size. From the top view, the MFR axis shows a significant
anti-clockwise rotation during the eruption. \Fig~\ref{eruption}c
shows a dimensionless current density, defined as $J\Delta/B$ (where
$J$ is the current density, $\Delta$ is the grid resolution and $B$ is
the magnetic field strength), on the central cross section, i.e., the
$x=0$ slice of the 3D volume. One can see a picture of the 2D standard
flare model: a plasmoid rises and leaves behind a cusp structure
corresponding to the edge of post-flare loop, and connecting them is a
long CS in which magnetic reconnection occurs continuously. This
reconnection results in high-speed bi-directional (up and down) plasma
jets due to the ``slingshot'' effect, and the upward jet flow
continuously pushes outward the newly reconnected magnetic
flux. Eventually, it produces a CME and drives an arc-shaped fast
magnetosonic shock enclosing the erupting structure.

\subsection{Evolution of Magnetic Squashing Factor and Twist Number}
To analyze the evolution of magnetic topology, in particular the
formation of MFR, during the simulated eruption, we calculate the
magnetic squashing factor $Q$ and magnetic twist number $T_w$ at the
bottom surface and at a vertical central cross section of the 3D
volume. The results are shown in \Fig~\ref{MFR_evol} for a few
snapshots and Supplementary Movie 2 for the whole evolution. At the
bottom surface, initially there are two thin strips of high $Q$, i.e.,
QSLs (or more precisely, the footprints of QSLs), forming two J shapes
on either side of the PIL. With onset of the eruption, the two
J-shaped QSLs on the bottom surface begin to evolve rapidly (see
\Fig~\ref{MFR_evol}a and the high-cadence evolution in Supplementary
Movie 2). In \Fig~\ref{recon_sample}a and b, we show the field lines
traced from the bottom QSLs at two different times, and the 3D
structure of the reconnection current sheet by an iso-surface of extremely strong current density.
As can be seen, all the field lines pass through the reconnection current sheet, which demonstrates
clearly that the field lines in the QSLs are undergoing reconnection. Consequently, the motion of the
bottom QSLs corresponds to the apparent motion of footpoints of the
field lines that were undergoing reconnection. Furthermore, from \Fig~\ref{recon_sample},
one can see that in the early phase the
reconnection is a fully 3D manner with a strong guide field component
(i.e., $B_x$) because joining in the reconnection is mainly the
strongly sheared, low-lying flux. While in the later phase, as the
sheared flux has been reconnected, the reconnection transfers into a quasi-2D
manner, which consumes mainly the large-scale, overlying flux that is
barely sheared.

On the central cross section (\Fig~\ref{MFR_evol}c),
the QSLs intersect with each other, developing into an X shape, i.e.,
an HFT~\citep{Titov2002}, and the intersection X point is essentially
the reconnection site (in analogy to the null point in a 2D X-shaped
reconnection configuration). As the eruption proceeds, more and more
magnetic fluxes reconnect, and consequently, the two J-shaped QSLs on
the bottom surface continuously separate with each other (see also
\Fig~\ref{recon_sample}c, in which the separation speed is
estimated). In the end of the simulation, they have swept to the
center of each magnetic polarity (which is analogous to the umbra of sunspot).
Meanwhile, the X point of the HFT rises upward
progressively (see also \Fig~\ref{recon_sample}d, in which the rising
speed is estimated) with the cusp region expanding below. Such two
QSLs with their separation should be manifested as two separating
flare ribbons in observations~\citep{Savcheva2016, JiangC2018}, while
the rise of the X point corresponds to the apparent rising of the apex
of post-flare arcades.

\Fig~\ref{MFR_evol}b and d present the $T_w$ distribution on the
bottom surface and the central cross section, respectively. Starting
from the hooks of the J-shaped QSLs, magnetic flux with high twist (as
denoted by large absolute values of $T_w$) begins to form owing to the
tether-cutting reconnection, which creates long field lines connecting
the far ends of the two pre-reconnection sheared arcades. With the
twisted flux accumulated through the continuation of reconnection, the
areas occupied by the footpoints of the highly twisted field lines at
the hooks expand. Consequently, the hook of each J-shaped QSL
continuously extends inward until its end reaches the arm, forming a
closed curve encircling the highly twisted flux (see the QSL at
$t=7$). Such a transition of QSLs reproduces the evolution of flare
ribbons that gradually forms close rings at their
ends~\citep{WangW2017}. With this, the MFR is {\it fully} separated
with its surroundings by the QSLs. Consistently, as can be seen in the
vertical cross section at $t=7$, the QSL form a closed tear-drop shape
connecting the HFT, within which the twisted flux of the MFR has twist
number $T_w$ mostly below $-2$.

The evolution speed of the QSLs is related to the rate of
reconnection. As shown in \Fig~\ref{recon_sample}c, at the beginning
of the eruption, the distance of the two QSLs is about $10$~Mm, and it
reaches $\sim 30$~Mm at the end of the simulation. The separation
speed first increases and then decreases, with its largest value of
about 18~km~s$^{-1}$ at the time of around $t=6.5$, which is also the
time the plasma acceleration reaches its maximum (see
\Fig~\ref{eruption}). Meanwhile, the rising speed of the X point,
i.e., the apex of the cusp structure reaches a maximum of about
40~km~s$^{-1}$. Our simulated flare-ribbon distances, their separation
speed as well as the rising speed of the cusp are all comparable to
typical observed ones~\citep{WangH2003, QiuJ2009SP, Hinterreiter2018,
  YanX2018}.

Nearly at the same time when the hook ends of the J-shaped QSLs close,
there is even a new QSL forms within the closed QSLs (see $t=10$ and
$t=13$ in \Fig~\ref{MFR_evol}a). In the positive polarity, for
example, this new QSL is bifurcated from the hook end and moves to the
right. As a result, the region bounded by the closed QSL, i.e., the
foot of the MFR, is divided into two regions separated by the newly formed QSL,
and the region after swept by the new QSL shows even stronger magnetic
twist than before. This indicates that there must be internal
reconnection between different field lines of the MFR. Another
noticeable change is the decrease of the area in the closed QSLs,
i.e., the feet of the MFR, which is quantified below.

\begin{figure}[htbp]
  \centering
  \includegraphics[width=0.45\textwidth]{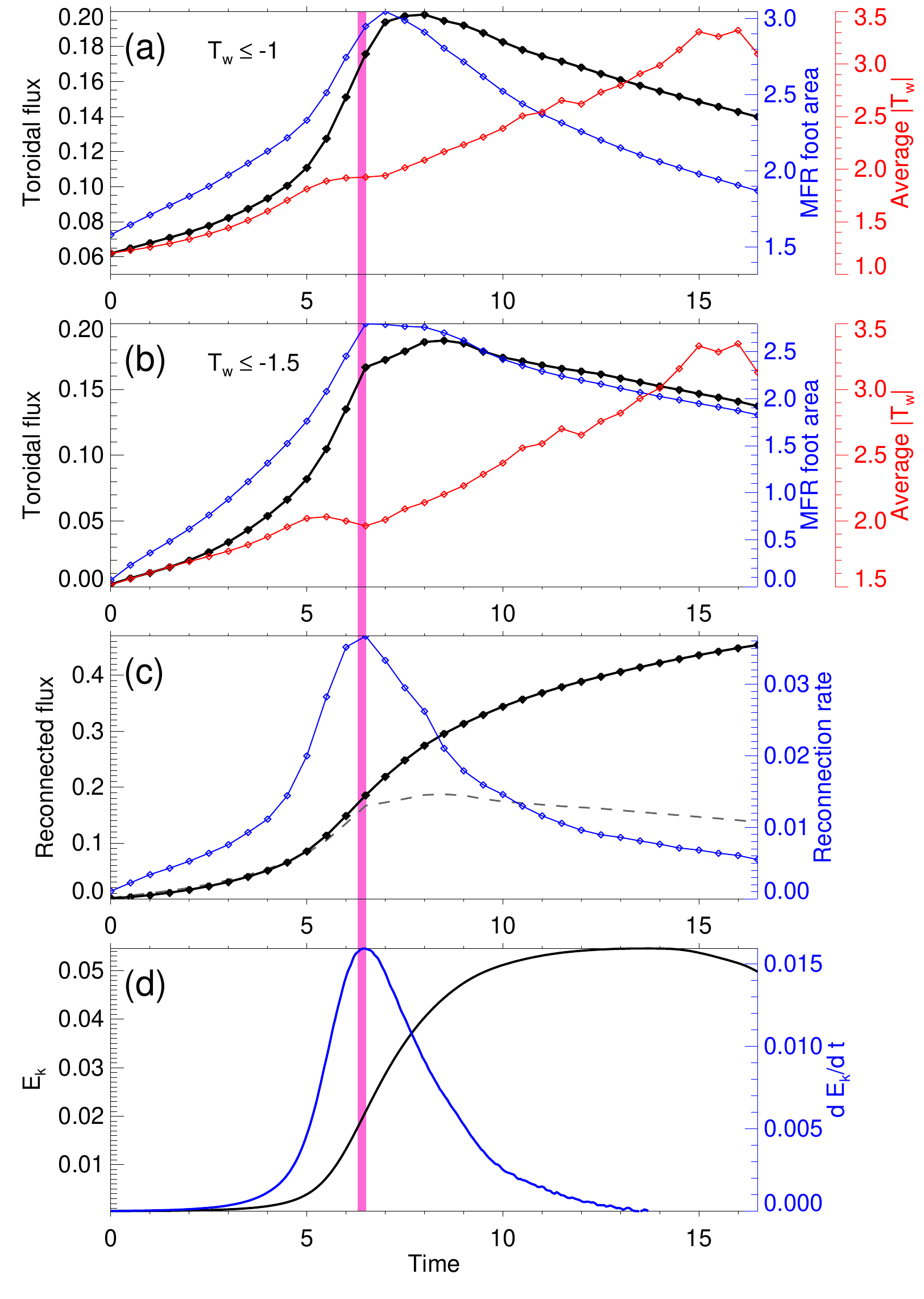}
  \caption{\textbf{Temporal evolution of different parameters in the
      eruption}. \textbf{(a)} Toroidal flux (black line), foot area
    (blue line) and average twist number (red line) of the MFR as
    calculated with $T_w \le -1$.  They are defined as, respectively,
    $\int_S B_z ds$, area of $S$ and
    $\int_S B_z T_w ds/\int_S B_z ds$, in which $S$ is the region of
    $T_w \le -1$ at the bottom surface.  \textbf{(b)} Same as
    \textbf{(a)} but with $T_w \le -1.5$. \textbf{(c)} Magnetic
    reconnected flux and its increasing rate. The dashed line shows
    the toroidal flux with $T_w \le -1.5$.  \textbf{(d)} Kinetic
    energy and its changing rate. The pink vertical bar denotes the
    peak time of the increasing rate of kinetic energy.}
  \label{MFR_recflux_evol}
\end{figure}

\begin{figure}[htbp]
  \centering
  \includegraphics[width=0.45\textwidth]{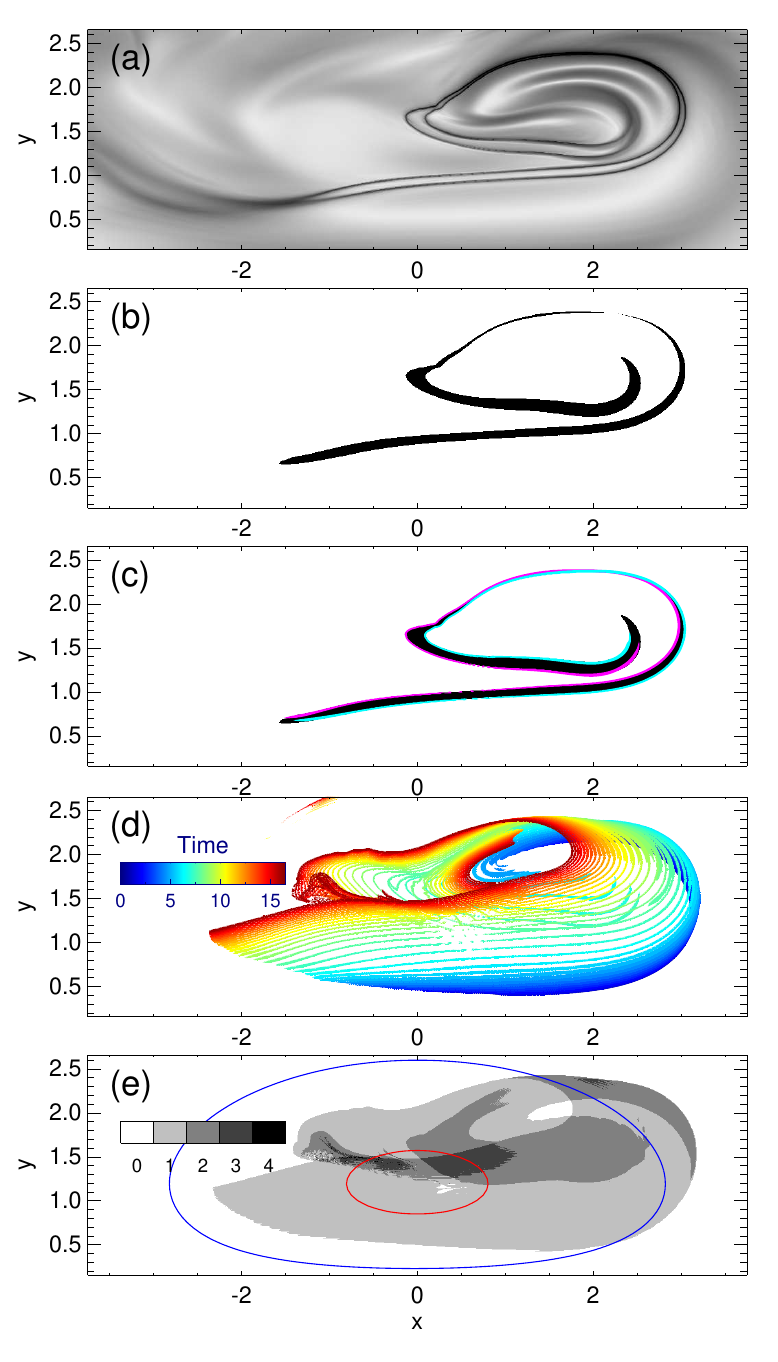}
  \caption{\textbf{Areas swept by the QSLs at the bottom surface}.
    \textbf{(a)} A overlaid image of $Q$ at two consecutive times
    $t=8$ and $8.5$.  \textbf{(b)} The reconnected region (shown in
    black) as calculated by using the slipping of the field line
    footpoints.  \textbf{(c)} Same as \textbf{(b)} and overlaid with
    the QSLs at $t=8$ (colored in cyan) and $t=8.5$ (colored in pink).
    \textbf{(d)} Overlaid plot of the QSLs at all the different times
    from $t=0$ to $16.5$. The QSLs at different times are color coded
    by time.  \textbf{(e)} All the regions swept by the QSL. Note that
    some regions are swept by the QSL with more than one time, and the
    swept times are denoted by gray color. Two contours of $B_z$, $10$
    (red) and $5$ (blue), are overlaid for showing the location of the
    magnetic polarity.}
  \label{recflux_method}
\end{figure}

\subsection{Evolution of MFR's Toroidal Flux}
The toroidal flux of the MFR, i.e., the content of the rope's flux
that connects the bottom surface, can be quantified by using the
distributions of twist number $T_w$ and squashing factor $Q$ at the
bottom surface. Before the full closing of the hook ends of the QSLs,
the feet of the MFR are characterized by the high $T_w$ areas, while
after the closing of the hooked QSLs, they can be identified more
accurately by the area within the closed QSLs, but the $T_w$ is still
a good indicator since the QSL-enclosed region has a distinctly strong
twist number (compare $T_w$ and $Q$ in \Fig~\ref{MFR_evol}). We thus
directly use the distribution of $T_w$ to locate the MFR's feet in the
whole evolution. However, it should be noted that the $T_w$ provides
only an approximation of the classic definition of the winding number
around a common axis, and that there is no consensus on the definition
of MFR based on either the winding number or the twist number $T_w$,
although it is generally agreed that the winding of field lines in an
MFR should be at least one turn. Therefore, we use two different
thresholds for $T_w$ to locate the MFR, and two values of the toroidal
flux of the MFR are calculated by summing the magnetic flux with $T_w$
exceeding the two thresholds, respectively. One is $T_w \le -1$, which
is also used by~\citet{DuanA2019} for searching MFRs in coronal field
extrapolations, and the other is $T_w \le -1.5$, which is properly
chosen such that the MFR can be clearly differentiated from the
background flux that has moderate twist number of $T_w \sim -1$ but
without reconnection during the eruption (thus remains non-flux-rope
field lines during the eruption). We also compute the areas of the MFR
foot using the two thresholds, as well as the average twist number of
the toroidal flux. The results are present in
\Fig~\ref{MFR_recflux_evol}a and b, which clearly show that the
toroidal flux (as computed by either thresholds) first increases,
reaching its peak value fast, and then decrease slowly. Such an
evolution pattern also applies to that of the MFR foot area. This
increase-to-decrease pattern of toroidal flux reproduces the observed
variations of magnetic flux in erupting MFR's foot as identified by
flare ribbons and transient coronal dimming~\citep{XingC2020}. On the
other hand, the mean twist number shows a systematic increase to a
value close to $3.5$ at the end of simulation.

\subsection{Evolution of Reconnection Flux}
We further quantify how much of the magnetic flux is reconnected
during the eruption. In principle, the total reconnected flux is
simply the flux (by a factor of two) that is swept by the QSLs at the
bottom surface in magnetic polarities of the same sign. This is
analogous to counting the photospheric magnetic flux swept by flare
ribbons to measure reconnection rate from direct
observations~\citep{QiuJ2002, WangH2003}. However, this requires a
very high time cadence of simulated data to capture the fast motion of
the QSLs, such that the combination of all the QSLs at different times
can seamlessly forms the whole area that experiences
reconnection. Furthermore, the geometry of QSLs is rather complex, and
it is not straightforward to compute the areas swept by the QSLs. For
instance, in \Fig~\ref{recflux_method}a, the QSLs at two consecutive
times ($t=6$ and $6.5$) are over-plotted. As can be seen, there are
clearly a margin between them, and this margin area is exactly the
region swept by the QSL in the time increment (i.e., from $t=6$ to
$6.5$). It is not easy to calculate the flux in this bounded area
owing to its very irregular shape.

Thus, we proposed an alternative way to calculate the reconnected flux
by taking advantage of the fast-slipping motion of the footpoints of
the reconnected field lines. In our simulation the bottom surface is
fixed without any motion, thus for any field line without
reconnection, it will be perfectly frozen with the plasma, and its two
footpoints will not change with time.  So, if tracing from a fixed
footpoint of a field line to the other end, the conjugate footpoint
will also be a fixed point at different times. If the field line
undergoes reconnection, the conjugate footpoint will slip to a
different location in the time step, and thus by the displacement one
can easily find whether the field line reconnects or not during the
time step. By this approach, the region between the two QSLs at the
two consecutive times is clearly enhanced, as shown in
\Fig~\ref{recflux_method}b and c, and then we can calculate the
reconnected flux in the time step. In \Fig~\ref{recflux_method}d all
the QSLs at different times (in one polarity) are overlaid, and in
\Fig~\ref{recflux_method}e all the regions swept by the moving QSL are
shown. We note that a large portion of the polarity is swept and
therefore reconnects during the eruption, and an evident drift of the
MFR foot can be seen. Interestingly, there are some regions that are
swept by the QSL with more than one time, some even reaching 4 times,
suggesting a rather complex internal reconnection in the erupting MFR.

The reconnection flux and its changing rate are shown in
\Fig~\ref{MFR_recflux_evol}c.  The total reconnected flux increases
monotonically, attaining nearly half of AR's total flux content at the
end of the simulation. The reconnection rate, i.e., the increasing
rate of the reconnected flux, shows an evolution pattern (i.e., fast
increase and then slow decrease) like the changing rate of the kinetic
and magnetic energies (see also \Fig~\ref{eruption}), and all of them
reach the peak at the same time. Such temporal correlation between
reconnection rate (or flare emission) and CME acceleration has been
well revealed in observation studies~\citep{ZhangJ2001, QiuJ2004},
stressing the central role and fundamental importance of magnetic
reconnection in producing flare and CME~\citep{ZhuC2020}.

It is interesting to compare evolution of the MFR's toroidal flux and
that of the reconnected flux. In \Fig~\ref{MFR_recflux_evol}c, the
dashed line shows the toroidal flux as present in
\Fig~\ref{MFR_recflux_evol}b. In the early stage, i.e., before the
reconnection rate reaches its peak, the reconnected flux almost equals
to the toroidal flux, meaning that the reconnection builds up the MFR
by transferring the same amount of sheared arcade flux into the same
amount of flux in the rope.  However, after the peak time, although
the reconnected flux continues to increase, the toroidal flux in the
rope decreases, and this suggesting that there must be reconnection
within the MFR, by the so-called \emph{rr-rf}
reconnection~\citep{Aulanier2019}. We note that, interestingly, the
peak time of reconnection rate (at $t=6.5$) also coincides with the
time of the closing of the QSLs, and immediately afterward, the
toroidal flux also reaches it maximum.

\section{Summary}
In this paper, we have studied the magnetic evolution of an MFR formed
during the eruption in an MHD simulation.  The MFR is generated
absolutely by tether-cutting reconnection of the pre-eruption,
strongly sheared arcade. In the early phase, the MFR is partially
separated from its ambient field by a QSL that has a double-J shaped
footprint on the bottom surface.  With the ongoing of the
reconnection, the arms (i.e., the straight parts) of the two J-shaped
footprints continually separate from each other, which eventually pass
through the centers of each polarity. Meanwhile, the hooks of the J
shaped footprints expand and eventually become closed almost at the
eruption peak time, and thereafter the MFR is fully separated with the
un-reconnected field by a QSL.  The reconnection substantially shapes
the MFR by first increasing quickly and then decreasing gradually its
total toroidal flux, which explains a recent observation of magnetic
flux variation in erupting MFR's foot.  In the whole eruption, nearly
half of the AR's flux is reconnected, and the reconnection rate, as
measured by the increasing rate of the reconnection flux, synchronizes
well with the energy conversion rate (i.e., magnetic energy releasing
rate and the kinetic energy increasing rate).  In the early stage,
i.e., before the reconnection rate reaches its peak, the reconnected
flux almost equals to the toroidal flux in the MFR, whereas after the
peak time the toroidal flux in the MFR decreases despite that the
reconnected flux continues to increase. The increase of toroidal flux
is owing to the flare reconnection in the early phase that transforms
the sheared arcade to twisted field lines, while its decrease should
be a result of reconnection between field lines in the interior of the
MFR in the later
phase, 
as first disclosed in \citep{Aulanier2019}.

Our simulation shows that the QSLs associated with the MFR in the
later phase become more complex than expected, since there are new
QSLs formed within the MFR, while the flux associated with these new
QSLs becomes extremely highly twisted. This is due to a fast expansion
of the MFR as well as its complex 3D nature, and thus at certain
locations field lines reconnect with others in the MFR or
themselves. Such reconnection may happen multiple times for field
lines rooted at the same locations, even making some of field lines
self-closed in the corona, which might be an important way for a CME
flux to be totally detached from the Sun.  The details of such
complexity and the involved reconnection, and whether such complexity
is hinted in observation, are to be elaborated in future works.

\acknowledgments This work is jointly supported by National Natural
Science Foundation of China (NSFC 4217040250, 41822404, 41731067) and
Shenzhen Technology Project JCYJ20180306171748011. The computational
work was carried out on TianHe-1(A), National Supercomputer Center in
Tianjin, China.


\clearpage

\end{document}